\documentclass[iicol]{sn-jnl}%Math and Physical Sciences Reference Style
\jyear{2021}%
\usepackage{graphicx}
\usepackage{amsmath}
\begin{document}

%\title[Dynamic Phases and Reentrant Hall Effect]{Dynamic Phases and Reentrant Hall Effect for Vortices and Skyrmions on Periodic Pinning Arrays}
\title{Dynamic Phases and Reentrant Hall Effect for Vortices and Skyrmions on Periodic Pinning Arrays}
\author*{\fnm{C. J. O.} \sur{Reichhardt}}\email{cjrx@lanl.gov}
\author{\fnm{C.} \sur{Reichhardt}}\email{reichhardt@lanl.gov}
%\equalcont{These authors contributed equally to this work.}
\affil*{\orgdiv{Theoretical Division}, \orgname{Los Alamos National Laboratory},
\orgaddress{\city{Los Alamos}, \postcode{87545}, \state{New Mexico}, \country{USA}}} 

\abstract{
We consider the motion of superconducting vortices and skyrmions on a square substrate near the first commensurate matching field. Slightly above commensuration, a series of dynamic phases appear including interstitial flow, and there is a transition from fluid flow to soliton flow that generates negative differential conductivity. Slightly below commensuration, vacancy depinning occurs. The dynamic phase transitions produce features in the velocity-force curves, differential mobility, and velocity fluctuations. When a Magnus force is also present, as in certain superconducting vortex or skyrmion systems, there is an expansion of the fluid state, and at lower drives there is a finite Hall angle in the fluid phase but a vanishing Hall angle in the soliton phase, giving rise to a reentrant Hall effect. We also find a regime where the Hall motion of the particles exhibits the same dynamic phases, including soliton motion at a finite angle that produces negative differential conductivity in the Hall response but not in the longitudinal response. Our results suggest that vortices driven over periodic pinning may be an ideal system for determining if a vortex Hall effect is occurring and would also be relevant for skyrmions at smaller Magnus forces. 
}


\maketitle

\section{Introduction}
Particles interacting with a substrate arise
in a variety of both
hard and soft condensed matter systems \cite{Bak82,Braun98,Reichhardt17}.
The ratio of the number of particles to the number of substrate
minima, known as the commensurability ratio or filling factor $f$,
is a useful measure when the substrate is periodic,
and at $f=1$
the system is said to be commensurate 
\cite{Bak82,Baert95,Harada96,Reichhardt98,Brunner02,Berdiyorov06,Reichhardt17}.
The substrate and particle arrangements
normally have the same symmetry
under commensurate 
conditions;
however, when the coupling to the substrate is weak, interactions among
the particles
can favor other kinds of ordering
\cite{Reichhardt17,Brazda18}. Under an applied drive, the
depinning force, friction,
and flow phases depend strongly on the
commensurability ratio
\cite{Bak82,Reichhardt97,Braun98,Bohlein12,Vanossi12,Vanossi13,Reichhardt17}. 
Just at commensuration, all the particle-particle interactions
cancel due to symmetry
and the depinning force $F_{c}$ reaches a maximum value.
On the other hand, if the
system is slightly above or below commensuration, solitons or kinks
can appear in the commensurate lattice that depin before the rest of
the particles,
strongly depressing the depinning threshold,
and in some cases multiple depinning transitions
can occur in which the depinning of kinks or anti-kinks
is followed by
depinning of the commensurate particles 
\cite{Reichhardt17,Bohlein12,Vanossi12}. 
For egg-carton or sinusoidal potentials,
which can be created optically, via lithographic pattering,
or through the periodic ordering of  atoms on a surface 
\cite{Bohlein12,Vanossi12,Hasnain13,Reichhardt17},
the kinks act like quasiparticles that are localized
above the commensurate lattice.
It is also possible
to create periodic
muffin tin potentials that have well-defined trapping
sites separated by extended flat regions
\cite{Baert95,Harada96,Reichhardt98}.  Such substrates have been realized for 
vortices in superconductors in the form of holes 
\cite{Baert95,Harada96} or magnetic dots \cite{Martin97}, as well as
for colloidal \cite{Mangold03} and skyrmion systems 
\cite{Reichhardt15a,Saha19}.

On either side of $f=1$ for
the muffin tin substrate,
incommensurations
take the form of
interstitial particles for $f>1$
or vacancies for $f<1$.
The interstitials appear
when all of the pinning sites are occupied and the additional particles
sit in
the flat part of the potential landscape between the occupied pinning sites. 
These interstitial particles can still
be pinned through their interactions with the particles
at the pinning sites, but exhibit distinct dynamical properties
compared to the directly pinned particles.
Reichhardt {\it et al.} \cite{Reichhardt97} showed that superconducting vortices
on such a muffin tin potential
exhibit multiple depinning transitions where the interstitial vortices
become mobile first, followed by the depinning of
vortices at the pinning sites, which interact with the
moving interstitials
to form a high mobility fluid phase.
Interestingly, at higher drives the system can 
organize into a soliton state with
one-dimensional (1D) flow along the 
pinning rows \cite{Reichhardt97,Reichhardt98}, causing the mobility to drop and
leading to negative differential conductivity. 
At higher drives,
all of the vortices can flow,
producing a jump in the average mobility. 
The motion observed in these different phases depends
on the pinning strength, density, and filling factor.
In superconducting vortex systems, the different phases
including negative differential resistivity have been found in several
experiments  \cite{Gutierrez09,Avci10}.
Other aspects have also been explored
such as adding asymmetry to the arrays \cite{Zhu01},
introducing local heating in order to produce both S-shaped and N-shaped
velocity-force curves in analogy to those found
in conduction curves for semiconductors \cite{Misko07},
and inducing density wave propagation \cite{daSilva11}. 
Numerous other studies of superconducting vortices in
periodic pinning arrays have revealed
multiple depinning transitions, the flow of interstitials,
kink flow, and multiple step jumps in the velocity-force curves
\cite{vanLook99,Reichhardt01b,Reichhardt08,Chen08,Pogosov10,Yetis11,Facio13,Chen14,Verga19}

Previous studies involved strictly overdamped particles;
however, non-dissipative forces can also arise, such as the
Magnus gyrotropic force \cite{Ao93,Sonin97}, which creates a velocity
component perpendicular to the forces acting on the particle
\cite{Ao93,Dorsey92}. In superconducting vortex systems, 
Magnus forces are possible,
and experiments have found evidence for 
transverse motion or the vortex Hall effect \cite{Lefebvre06},
with more recent observations showing
vortex Hall angles of up to $45^\circ$ 
in certain types of superconducting systems \cite{Ogawa21}.
For vortices in superfluids
with pinning, the Magnus force can strongly affect the dynamics
\cite{Wlazlowski16}.
It is possible to create experimental realizations of
vortices
in superfluids with periodic pinning \cite{Tung06}, and there have been 
numerous studies of dynamical phases of vortices in
Bose-Einstein condensates with periodic
pinning where the Magnus force is relevant \cite{Kasamatsu06}.
There are also other particle-like systems
with strong Magnus forces that could
be investigated with periodic substrates,
such as active spinner and chiral active matter systems
\cite{Reichhardt20b,Banerjee17,Bililign22}. 

Another particle-like system that can interact
with a periodic pinning array is skyrmions, which are
a magnetic texture that can arise in chiral magnets
\cite{Yu10,Nagaosa13,Muhlbauer09}. Magnus forces can be quite
significant in skyrmion systems
\cite{Nagaosa13,EverschorSitte14} 
and experiments have directly demonstrated
what are called skyrmion Hall angles as large as
$30^\circ$ to $45^\circ$ \cite{Jiang17,Litzius17}, while in
other materials the skyrmion Hall angles are smaller but nonzero
\cite{Woo18,Reichhardt20,Zeissler20}.
It is known that the skyrmion Hall angle is modified when the
skyrmions interact with pinning,
adopting a small value for low skyrmion velocities just above depinning
and increasing with increasing skyrmion velocity until
the skyrmion Hall angle approaches
its intrinsic value at high drives
\cite{Reichhardt15,Reichhardt15a,Kim17,Jiang17,Litzius17,Legrand17,Juge19,Zeissler20}.
For skyrmions interacting with periodic pinning,
commensuration effects can arise, and
the Magnus  
force produces a symmetry locking effect in which the
skyrmions preferentially move along 
certain symmetry directions of the pinning lattice,
causing the skyrmion Hall angle to be quantized
\cite{Reichhardt15,Reichhardt18,Feilhauer20}. Simulations of skyrmions
with large
Magnus forces indicate that clustering effects can
also occur  \cite{Reichhardt18}.
Experimentally, periodic substrates for skyrmions
can be produced using nanoscale arrays \cite{Saha19,Juge21,Kern22} or with
optical trapping. 

In this work we reexamine the dynamic phases for
superconducting vortices interacting with a periodic pinning array 
for filling factors ranging from $f = 0$ up to $f = 2.3$ in the
absence and presence of a Magnus force.
We highlight the flow of vacancies and
interstitials,
identify fluid and soliton flow regimes, and show the
changes in the velocity-force curves, including negative
differential conductivity.
We also consider the changes in the fluctuations across the transitions,
which in some cases more clearly reveal
the boundaries between the different dynamical states. 
With this background in place, we examine the effect of
adding a finite Magnus force, which is quantified by the intrinsic Hall angle
$\theta_{H}^{\rm int}$. 
For Hall angles of
$\theta_{H}^{\rm int}=10^\circ$ or less,
the dynamic phases are similar to those found
in the overdamped system except that the
fluid phase has a finite Hall angle, and 
an additional phase
emerges at higher drives where the bulk vortices move at a 
finite Hall angle while
interstitials, vacancies, and solitons move without a Hall angle.
For higher intrinsic Hall angles,
there is a new set of phases
that show negative differential conductivity in the Hall velocity but not in the
longitudinal velocity when
a transition occurs from a fluid phase to
a phase in which
solitons move at an angle.
Our results could be useful in determining if certain
superconducting vortex systems have a finite Magnus force,
and are also relevant for
skyrmion systems with small intrinsic skyrmion Hall angles.
Additionally, our work suggests that
in skyrmion systems, the flow of kinks, antikinks,
or interstitials could be used to lower or mitigate the
skyrmion Hall effect.  

\begin{figure}
\centering
\includegraphics[width=\columnwidth]{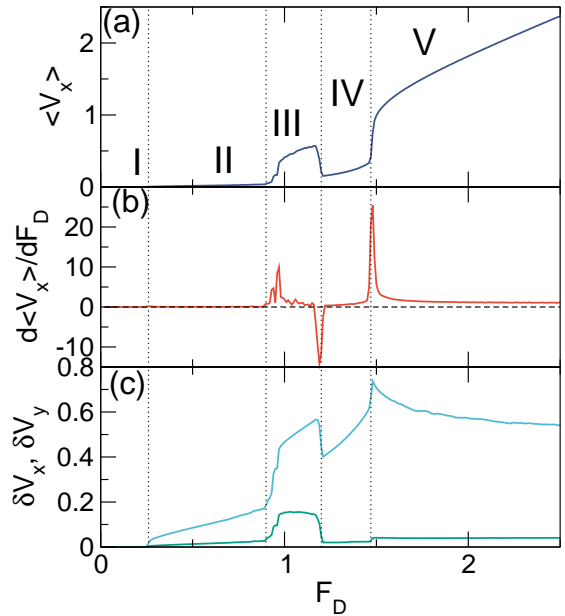}
\caption{
  (a) The velocity-force curve
$\langle V_x\rangle$ vs $F_{D}$ 
for a system with
a square pinning array
at a filling of $f = 1.0368$
for $\alpha_{m} = 0.0$, $\alpha_d=1.0$, and $F_{p} = 1.5$.
The five phases  are: I (pinned), II (interstitial flow), 
III (disordered liquid), IV (soliton motion), and V (moving smectic).
(b) The corresponding $d\langle V_{x}\rangle/dF_{D}$ vs $F_D$
showing negative differential conductivity at the III-IV transition. 
(c) $\delta V_{x}$ (blue) and $\delta V_y$ (green) vs $F_D$.
}
\label{fig:1}
\end{figure}

\section{Model}   

We simulate an $L \times L$
system with periodic boundary conditions in the $x$ and $y$ directions.
The sample contains
a square array of 
$N_{p}$ pinning sites
modeled as localized trapping sites of finite 
radius $r_p$ with a lattice constant $a_{p}$, where
$a_{p} > r_{p}$ so that the system is in a muffin tin potential regime. 
We add
$N_v$ superconducting vortices to the sample, and the system is characterized 
by a filling factor $f  = N_{v}/N_{p}$, where at $f = 1.0$ there is one vortex
per pinning site.
When the pinning sites are small, as we consider here, only one
vortex can be captured by each
pinning site, and when $f > 1.0$,
the additional vortices are located in the interstitial regions 
between the pinned vortices.
Commensuration effects appear for integer $f=1$, 2, 3...
when the interstitial vortices form ordered lattices
at integer fillings
and disordered configurations at non-integer fillings
\cite{Reichhardt98}. Similar effects occur
for colloidal particles
or skyrmions on muffin tin potentials \cite{McDermott13a,Reichhardt18}.
When $f < 1.0$, a well defined number of
vacancies or holes appear in the commensurate pinned lattice. 
In this work we focus on the range $0 < f < 2.5$, with
particular emphasis on the region near $f = 1.0$.
The dynamics of the particles 
are obtained by integrating the following equation of motion: 
\begin{equation}
\alpha_{d}{\bf v}_{i} + \alpha_{m}{\hat z}\times {\bf v}_{i} = {\bf F}^{vv}_{i} + {\bf F}^{sp}_{i} + {\bf F}_{D} .
\end{equation}
Here
${\bf r}_{i}$ is the location of vortex $i$,
${\bf v}_{i} = d{\bf r}_{i}/dt$ is the 
net vortex velocity,
and $\alpha_{d}$ is the damping term
that aligns the velocity in the direction of
the net forces on the vortex. The cross term is
the Magnus force with prefactor
$\alpha_{m}$ that generates a
velocity component perpendicular to the net forces on the vortex.
Previous vortex simulations treated the regime
where there is only damping  \cite{Reichhardt98}
and the Magnus force term is absent.

The repulsive vortex-vortex interactions have the form
${\bf F}_i^{vv}=\sum_{j\neq i}^{N_v}K_{1}(r_{ij}){\bf \hat r}$,
where $r_{ij} = \lvert{\bf r}_{i} - {\bf r}_{j}\rvert$ is the distance between
vortex $i$ and vortex $j$
and ${\bf \hat r}=({\bf r}_i -{\bf r}_j)/r_{ij}$.
Here $K_1$ is the modified Bessel function
that falls off exponentially for large $r_{ij}$. 
The repulsive Bessel function interaction applies to both vortex-vortex and
skyrmion-skyrmion interactions \cite{Lin13}.
Since the interaction falls off
rapidly for $r_{ij} > 1.0$, we place a cutoff on the interactions 
for $r_{ij}> 7.0$ for computational efficiency.  
The pinning sites are modeled as parabolic potential traps
with a maximum pinning force of $F_{p}$.
The driving term representing the Lorentz
force from an applied current is given by
${\bf F}_D=F_D{\bf \hat{x}}$.
An isolated vortex in
a pinning site only becomes depinned when $F_{D}/F_{p} > 1.0$. 

\begin{figure}
\includegraphics[width=\columnwidth]{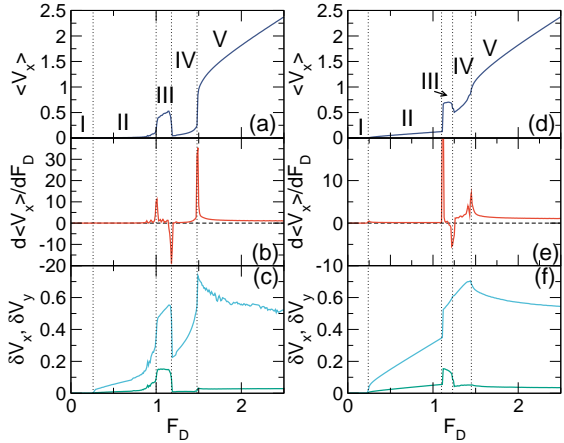}
\caption{
Transport signatures
for a system
with
$\alpha_m=0$, $\alpha_d=1.0$, and $F_p=1.5$
at (a,b,c) $f=1.005$ and (d,e,f) $f=1.1392$.
The five phases  are: I (pinned), II (interstitial flow), 
III (disordered liquid), IV (soliton motion), and V (moving smectic).
(a,d) $\langle V_x\rangle$ vs $F_{D}$.
(b,e) The corresponding $d\langle V_{x}\rangle/dF_{D}$ vs $F_D$ curves
contain regions of negative differential conductivity.
(c,f) $\delta V_{x}$ (blue) and $\delta V_y$ (green) vs $F_D$.
}
\label{fig:2}
\end{figure}

In this work, we fix $L = 36$,
$N_{p} = 625$, $a = 1.44$, and $r_{p} = 0.25$.
We select $\alpha_m$ and $\alpha_d$ according to the relation
$\alpha_m^2+\alpha_d^2=1$ in order to facilitate comparisons
of velocity-force curves among systems with different Magnus
terms.
We
perform simulated annealing by
initializing the system in a high temperature state and
lowering the temperature to $T = 0.0$, after which we
apply a drive in increments
of $\Delta F_{D}=0.01$
and average the velocities
over a period of $2.6 \times 10^5$ simulation time steps
before advancing to the next drive increment. 
In previous work \cite{Reichhardt98}, the amount of averaging time per step was
much smaller, 
which gave highly fluctuating velocity signals in the disordered
flow regimes.
With the much longer time averaging employed here,
the velocity-force curves are smoother.
For each drive increment, we measure the average vortex velocity
$\langle V_{x}\rangle = \sum^{N_v}_i{\bf v_i}\cdot {\hat {\bf x}}$ and 
$\langle V_{y}\rangle = \sum^{N_v}_i{\bf v_i}\cdot {\hat {\bf y}}$. 
We also measure the standard deviation of the vortex velocities in
the $x$ and $y$ directions,
$\delta V_{x} = \sqrt{[\sum^{N_v}_{i}({\bf v_i}\cdot {\hat {\bf x}})^2
    - \langle V_{x}\rangle^2]/N_v}$ and
$\delta V_{y} = \sqrt{[\sum^{N_v}_{i}({\bf v_i}\cdot {\hat {\bf y}})^2 - \langle V_{y}\rangle^2]/N_v}$. 
The Hall angle is defined as
$\theta_{H} = \arctan(\langle V_{y}\rangle/\langle V_{x}\rangle)$. 
In the absence of quenched disorder,
the intrinsic Hall angle is $\theta_{H}^{\rm int} = \arctan(\alpha_m/\alpha_d)$.

\begin{figure}
\includegraphics[width=\columnwidth]{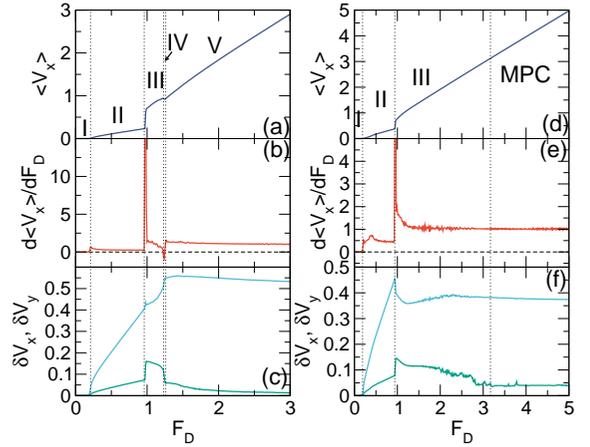}
\caption{Transport signatures for a system
with
$\alpha_m=0$, $\alpha_d=1$, and $F_p=1.5$ at
(a,b,c) $f=1.34$ and (d,e,f) $f=1.76$.
The five phases are: I (pinned), II (interstitial flow), III (disordered
liquid),
IV (soliton motion), and V (moving smectic).  
(a,d) $\langle V_x\rangle$ vs $F_D$.
(b,e) The corresponding $d\langle V_{x}\rangle/dF_{D}$ curves,
where negative differential conductivity is present only for $f=1.34$.
(c,f) $\delta V_{x}$ (blue) and $\delta V_{y}$ (green) vs $F_D$.
At $f=1.34$, phase IV is extremely narrow. At $f=1.76$,
phases IV and V are absent
but a moving polycrystalline phase (MPC) appears
at higher drives.
}
\label{fig:3}
\end{figure} 

\section{Zero Hall Effect}

\begin{figure*}
\includegraphics[width=0.75\textwidth]{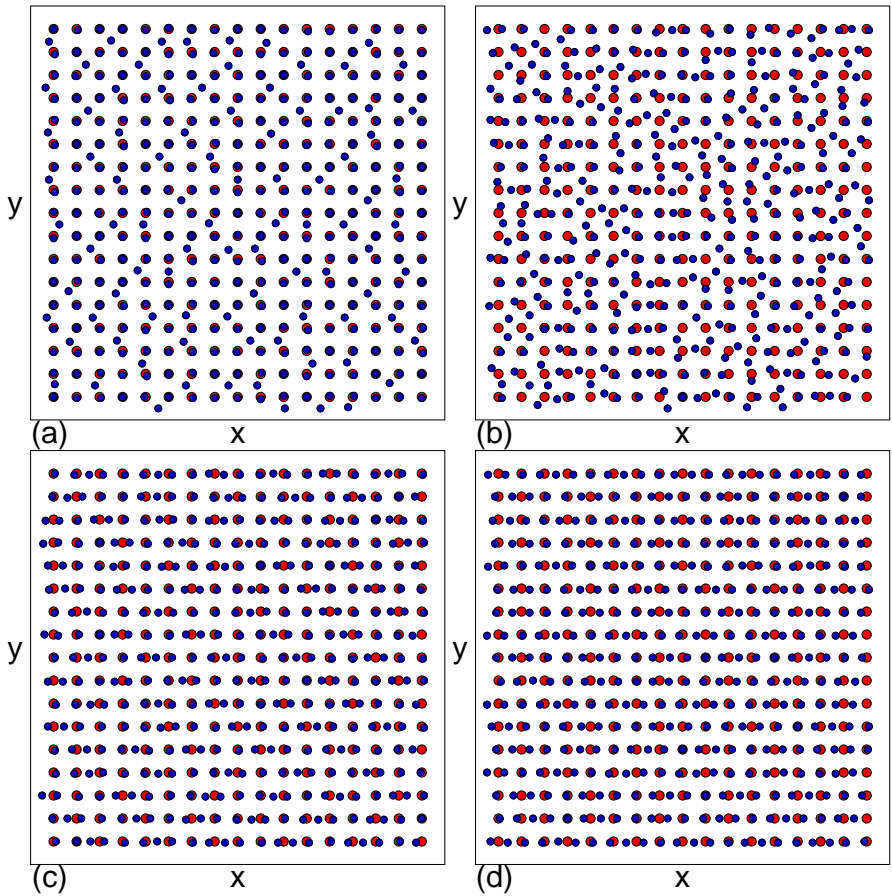}
\caption
{Images of vortex (blue) and pin (red) positions
in a portion of the system in
Fig.~\ref{fig:3}(a,b,c) with $\alpha_m=0$, $\alpha_d=1$,
$F_p=1.5$, and $f=1.34$.  
(a) Pinned phase I
at $F_{D} = 0.1$.
(b) Disordered liquid flow in phase III.
(c) Soliton motion along 1D rows in phase IV.
(d) Phase V or the moving smectic phase. 
}
\label{fig:4}
\end{figure*}

We first consider the zero intrinsic Hall angle situation with
$\alpha_{m} = 0.0$
and $\alpha_{d} = 1.0$.
In Fig.~\ref{fig:1}(a,b) we plot $\langle V_{x}\rangle$ and
$d\langle V_x\rangle/dF_D$
versus $F_{D}$  
for a system with $F_{p} = 1.5$ and $f = 1.0368$,
just above the first matching filling.
We identify the same five distinct dynamical phases found previously
\cite{Reichhardt18,Zhu01,Misko07}, consisting of
a pinned phase I for
$F_{D} < 0.26$, 
a sliding interstitial phase II where the small number of vortices
that are not in the pinning sites move
for $0.26 \leq F_{D} < 0.9$,
a disordered fluid flow phase III for
$0.9 \leq F_{D} < 1.2$, a soliton flow phase IV for $1.2\leq F_D<1.47$, and
a moving smectic phase V for $F_D \geq 1.47$.
There is an upward spike in
$d\langle V_{x}\rangle/dF_{D}$ at the II-III boundary
when a large number of vortices suddenly begin to move.
The transition to phase IV where all the vortices are moving is
accompanied
by a sharp drop in the velocity,
leading to a region of negative differential
conductivity  with
$d\langle V_{x}\rangle/dF_{D} < 0.0$.
A peak in $d\langle V_{x}\rangle/dF_{D}$ appears at the IV-V transition.
In Fig.~\ref{fig:1}(c), the $\delta V_x$ and $\delta V_y$ versus $F_D$ curves
provide a clearer signature of phase II followed by a large upward
jump into phase III where the flow is more disordered.
At the transition to phase IV, both $\delta V_x$ and $\delta V_y$ jump
down 
when the motion becomes 1D again,
while the IV-V transition is marked by a
cusp in $\delta V_{x}$ and a smaller jump in $\delta V_{y}$.
We note that even though $\langle V_{y}\rangle$ 
is zero for all drives in the overdamped limit,
$\delta V_{y}$ is always finite above the depinning transition. 
The difference between $\delta V_x$ and $\delta V_y$ reaches its
largest value 
at the IV-V transition
when a large number of pinned vortices coexist with a smaller number of
vortices that are flowing along 1D channels.
Within the moving smectic phase V, 
$\delta V_{x}$ remains large since all
of the vortices are moving but rows that are more highly occupied
travel faster than the other rows, giving variability in the $x$ velocity
that gradually diminishes with increasing drive.
There is a small jump in $\delta V_y$ at the IV-V transition due to
a slight increase in the ability of the vortices to wiggle in the direction
transverse to the drive once all the vortices have depinned.
In previous work on a system of this type \cite{Reichhardt18,Zhu01,Misko07},
the $d\langle V_x\rangle/dF_{D}$ curves were not measured since long
simulation times are required in order to obtain well-averaged data that
can be differentiated; 
additionally, neither $\delta V_{x}$ nor $\delta V_{y}$ were  measured. 

In Fig.~\ref{fig:2}(a,b,c) we show
transport curves for the same system
at a lower filling of $f = 1.005$ where the
number of interstitial vortices is smaller.
The III-IV transition becomes even
sharper, as indicated by $\delta V_{x}$  and
$\delta V_{y}$ versus $F_D$ in Fig.~\ref{fig:2}(c).
Even though the number of moving interstitial vortices is reduced,
these vortices still create an instability among the pinned vortices 
that leads to the emergence of the liquid phase III,
while in phase IV the motion
is strictly 1D and the smaller number
of interstitials are strongly localized in the
form of incommensurations
sliding along the rows of pinning sites.
Figure~\ref{fig:2}(d,e,f) shows that the same phases occur at a higher
filling of
$f = 1.1392$,
but the downward jump in
$\langle V_{x}\rangle$ at the III-IV transition is reduced in
size since a larger number of solitons are present in phase IV.
At this filling, where the number of interstitial vortices is
approximately 14\% of the number of pinning sites,
the soliton is no longer localized at a single vortex
but consists of a group of approximately three
moving vortices that are displaced from the pinning sites, while
the total number of moving vortices in phase III is close to 50\%.
There is no drop in $\delta V_x$ at the III-IV transition for this filling,
but the drop in $\delta V_{y}$ at the transition to phase IV persists.
A cusp still appears in $\delta V_{x}$
at the IV-V transition.

In Fig.~\ref{fig:3}(a,b,c), the transport curves
for the same system at 
$f = 1.34$
contain only a very small window of phase IV
that appears
as a small region of negative differential conductivity in
Fig.~\ref{fig:3}(b). 
Since each soliton pulse involves about three vortices,
for $f > 1.34$ almost all of the vortices
are moving in phase III, and there is no longer
a dip in $\langle V_x\rangle$ at the III-IV transition. 
The I-II, II-III, and IV-V transitions
produce clear signatures in
$\delta V_{x}$ and $\delta V_y$. 

At $f=1.76$ in Fig.~\ref{fig:3}(d,e,f),
the transport curves indicate that phases IV and V are lost.
There is still a pinned phase I and a transition 
to the interstitial flow phase II.
Across the II-III transition, there are
peaks in $d\langle V_{x}\rangle/dF_{D}$, $\delta V_{x}$, and $\delta V_y$;
however, there are now
enough interstitial vortices present that
confinement of the flow
strictly along the pinning rows
in phase IV is too energetically costly to occur,
so phase III becomes greatly extended.
When $F_{D}> 3.0$,
the drive is large enough to overcome the effectiveness of the pinning,
and the vortices
partially reorder into a moving
polycrystalline (MPC) state where some vortices move along the
pinning rows while others do not. 
There is no signature of the III-MPC transition in
$\langle V_{x}\rangle$ or $d\langle V_{x}\rangle/dF_{D}$, but
it does appear as a drop in $\delta V_{y}$. 

\begin{figure}
\includegraphics[width=\columnwidth]{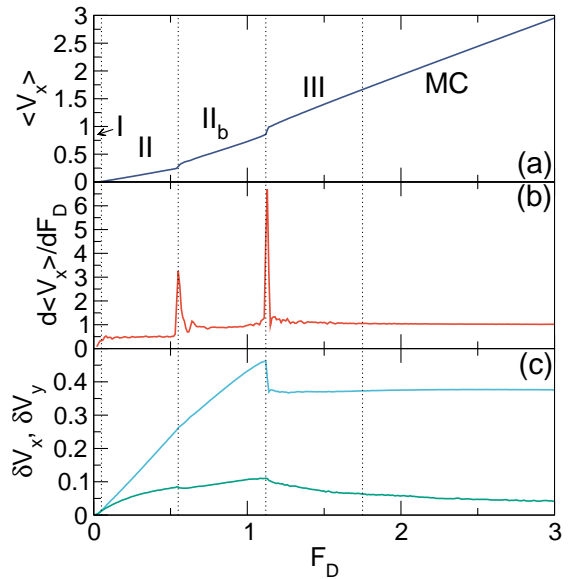}
\caption{(a)
$\langle V_x\rangle$ vs $F_{D}$
for a system 
with $\alpha_m=0$, $\alpha_d=1$, and $F_p=1.5$ at
$f = 2.5$.
The phases are: I (pinned), II (interstitial flow),
II$_b$ (combination of phases II and IV),
III (disordered liquid), and MC (moving crystal).
(b) The corresponding $d\langle V_{x}\rangle/dF_{D}$
vs $F_D$ showing a pair of peaks marking the II-II$_b$ and II$_b$-III
transitions.
(c) $\delta V_x$ (blue) and $\delta V_y$ (green) vs $F_D$. 
}
\label{fig:5}
\end{figure}

\begin{figure*}
\includegraphics[width=0.75\textwidth]{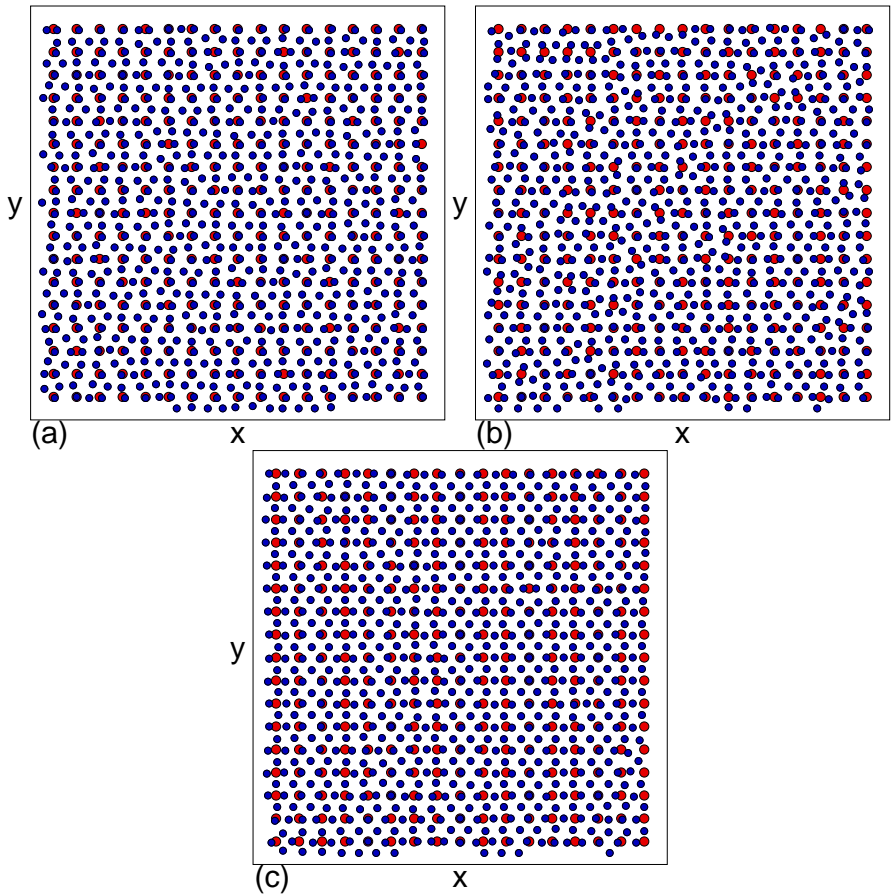}
\caption
{Images of vortex (blue) and pin (red) positions in a portion
of the system from
Fig.~\ref{fig:5} with $\alpha_m=0$, $\alpha_d=1$, $F_p=1.5$, and $f=2.5$.
(a) Phase II$_b$.
(b) Disordered liquid flow in phase III.
(c) The moving crystal phase MC.  
}
\label{fig:6}
\end{figure*}

In Fig.~\ref{fig:4}(a) we illustrate
the pinned phase I for the system in Fig.~\ref{fig:3}(a,b,c) with
$f = 1.34$.
Here all of the pinning sites are filled and the additional vortices sit in
the interstitial regions.
During phase II flow,
these interstitial vortices move while
the other vortices remain pinned, and the configuration looks like a
translating version of Fig.~\ref{fig:4}(a).
In phase III, shown in Fig.~\ref{fig:4}(b),
the vortices are disordered and
move in both the $x$ and $y$ directions, leading to the
increase in both $\delta V_{x}$ and $\delta V_{y}$ found in Fig.~\ref{fig:3}(c).
Figure~\ref{fig:4}(c) indicates that in phase IV,
the vortices form a series of 1D channels
with
density modulations
at the locations of the moving solitons.
In phase V, illustrated in Fig.~\ref{fig:4}(d),
all of the vortices are moving and there is
smectic ordering.
The vortex density along each 1D row is more uniform in phase V than
in phase IV.
Phase V occurs only
for $f < 1.5$. For higher fillings, at higher drives the system can 
form a moving polycrystalline or crystalline phase.

For $2.0 < f < 2.5$, the square checkerboard state
that appears at $f = 2.0$ contains interstitials.
These interstitials depin first at a low depinning force
and the system enters a state
that we call phase II$_a$ in which solitons flow through
the interstitial lattice.
For $2.5 \leq f < 3$, there are enough interstitials present that all of
the interstitials move simultaneously above the depinning
threshold and adjacent rows of vortices sitting on the pinning sites
slide with respect to each other to produce what we call
phase II$_b$, which is a combination of phase II and phase IV.  

In Fig.~\ref{fig:5}
we show $\langle V_{x}\rangle$,
$d\langle V_{x}\rangle/dF_{D}$, 
$\delta V_{x}$ and $\delta V_y$ versus $F_D$ for a
system
at $f = 2.5$.
There are two prominent peaks in $d\langle V_x\rangle/dF_D$ at the
II-II$_{b}$ and II$_b$-III transitions.
In Fig.~\ref{fig:6}(a) we illustrate the
vortex positions in phase II$_b$ for the system in
Fig.~\ref{fig:5}. Figure~\ref{fig:6}(b) shows
the disordered flow in phase III, while
in Fig.~\ref{fig:6}(c), in the moving crystal (MC)
state, the vortices form a
large scale crystal with a uniform orientation.

\begin{figure}
\includegraphics[width=\columnwidth]{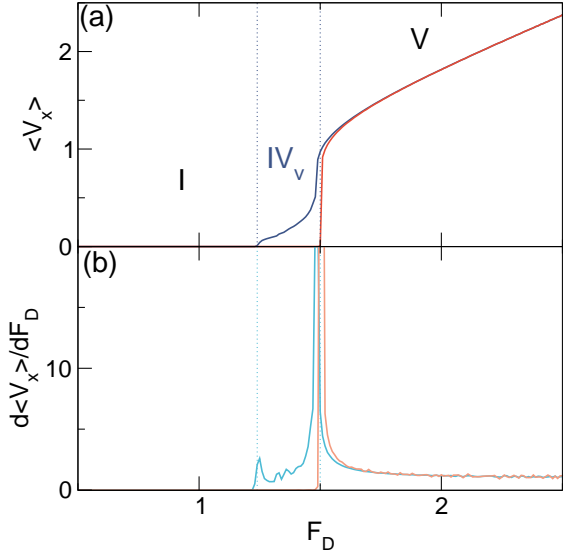}
\caption
{(a) $\langle V_x\rangle$ vs $F_D$ for a system 
with $\alpha_m=0$, $\alpha_d=1$, and $F_p=1.5$ at
$f = 0.9536$ (blue) and $f  = 1.0$ (red).
(b) $d\langle V_{x}\rangle/dF_{D}$ for the same system at
$f=0.9536$ (light blue), where there are two peaks,
and $f=1.0$ (orange), where there is only one peak.
The phases are: I (pinned), IV$_v$ (vacancy soliton motion for
$f=0.9536$ sample only), 
and V (moving smectic).
}
\label{fig:7}
\end{figure}

\begin{figure}
\includegraphics[width=\columnwidth]{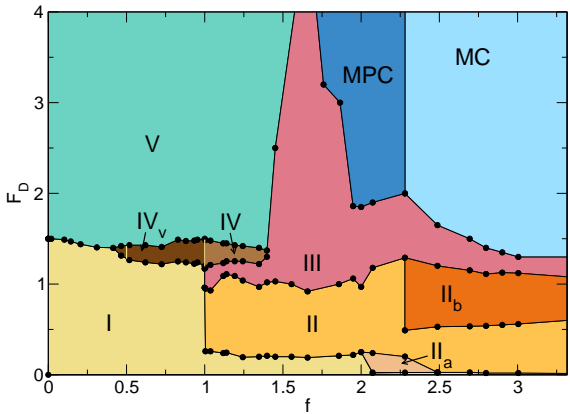}
\caption
{Dynamic phase diagram as a function of $F_D$ vs $f$ for 
systems with $\alpha_m=0$, $\alpha_d=1$, and
$F_p=1.5$  
constructed from the  $\langle V_x\rangle$,
$d\langle V_x\rangle/dF_{D}$,
$\delta V_{x}$, and $\delta V_{y}$ curves,
showing phase I (pinned, yellow),
phase II (interstitial flow, light orange),
phase II$_a$ (soliton flow in the interstitial lattice, light brown),
phase II$_b$ (interstitial motion and flow along the pinning rows,
dark orange),
phase III (disordered flow, pink),
phase IV (soliton flow, medium brown),
phase IV$_{v}$ (vacancy flow, dark brown),
phase V (moving smectic, green),
MPC (moving polycrystalline, medium blue),
and MC (moving crystal, light blue).}
\label{fig:8}
\end{figure}

For $0.5 < f < 1.0$, only three phases
are present: the pinned phase, a moving vacancy or anti-kink phase
termed IV$_v$,
and the moving smectic phase.
In Fig.~\ref{fig:7} we plot $\langle V_{x}\rangle$ and
$d\langle V_{x}\rangle/dF_{D}$
for a system
with $f = 0.9536$.
The ground state at this filling
consists of a square lattice containing some vacancies.
At $F_{D} = 1.24$, 
these vacancies depin first and the system enters phase IV$_v$,
which coincides with
the first peak in the differential conductivity curve.
The remaining vortices depin
near $F_{D} = 1.5$,
producing
the second peak in the differential conductivity curve.
Also appearing in the figure is the behavior
for a sample with $f=1.0$,
where there is a single depinning transition from phase I directly into
phase V at $F_D=F_p=1.5$.
A similar single depinning transition appears
when the number of vacancies becomes very large for
$f < 0.5$.
At higher drives for $0.5 < f < 1.0$,
all of the vortices flow along the pinning rows and the behavior
resembles phase V motion but with moving antikinks instead of
moving kinks.

From the features in the velocity-force,
$d\langle V_x\rangle/dF_{D}$, $\delta V_{x}$, and $\delta V_y$ curves,
we construct a diagram of the different phases, shown
in Fig.~\ref{fig:8}, as a function of $F_D$ versus $f$.
The pinned phase I has a peak at $f=0.5$ and a larger peak
precisely at $f=1.0$. At these peaks, and at low $f$, the system
depins directly from phase I to phase V smectic flow.
For $0.5 < f < 1.0$, there is a two step depinning transition, with the
system first depinning
from phase I to the vacancy or antikink
flow state in phase IV$_v$,
and then depinning into phase V at higher drives.
For $1.0 < f < 1.4$, the
system depins from phase I into the interstitial flow phase II,
followed by a transition associated with negative differential conductivity
into the disordered flow phase III.
At higher drives
in this range of fillings, there is a transition to phase IV
with soliton flow
along the pinning rows, followed at high drives by phase V.

For $1.4 < f < 1.65$ in Fig.~\ref{fig:8},
the disordered phase III flow increases significantly in extent and
extends up to very high drives.
This is caused by a competition between the pinning energy, which favors having
all of the vortices flow along the pinning rows as in phase IV, and the
vortex-vortex interaction energy, since at these higher densities,
the vortex density along a given pinning row would become prohibitively
high if all of the
vortices were accommodated along the pinning rows. What happens
instead is that vortices along the pinning rows buckle and relieve
the vortex-vortex interaction energy by ejecting some vortices into the
interstitial region. These vortices then attempt to fall back into the pinning
rows by ejecting other vortices,
and the result is a highly disordered flow state.
For $1.65 < f <  2.25$,
at high drives the vortices
partially organize into the polycrystalline state described
in Fig.~\ref{fig:1},
while for $f \geq 2.25$, an ordered moving crystal (MC) state
can form.
The MC emerges when the filling becomes high enough that
the buckling of vortices out of the pinning rows produces rows of interstitial
vortices that have roughly the same density as the rows of vortices
moving along the pinning sites. This stabilizes the interstitial rows and
puts a halt to the ejection process found at lower fillings.

For $2.0 < f < 2.5$ in Fig.~\ref{fig:8}, the initial depinning
is followed by a small window
of phase II$_a$ in which only the solitons in the
interstitial lattice move before the system enters the interstitial
phsae II flow.
For $f > 2.25$, a new phase emerges above phase II that we call phase
II$_b$, where the interstitial vortices are flowing 
but there is also flow of some of the vortices along the
pinning rows. The II$_b$ flow is generally 1D in nature, and is followed
by a transition
to the disordered fluid phase III
and then by a transition into a moving crystal at higher drives.

The phase diagram shown in Fig.~\ref{fig:8} has several features that
differ from the phase diagram obtained in earlier work
\cite{Reichhardt98} due to the higher pinning density we
consider here.
Additionally, in Ref.~\cite{Reichhardt98}, the phase diagram
boundaries were determined using only the velocity-force curves,
but here we add information from $d\langle V_x\rangle/dF_{D}$,
$\delta V_{x}$, and $\delta V_y$ to better characterize all of the phases.
The biggest difference is that 
a large window of the disordered phase III appears
around $f = 1.6$ in Fig.~\ref{fig:8},
since the denser pinning brings vortices in the interstitial regions
closer to the vortices flowing along the pinning rows and enhances the
ejection process that leads to the disordered flow.

\section{Finite Magnus Force}

\begin{figure}
\includegraphics[width=\columnwidth]{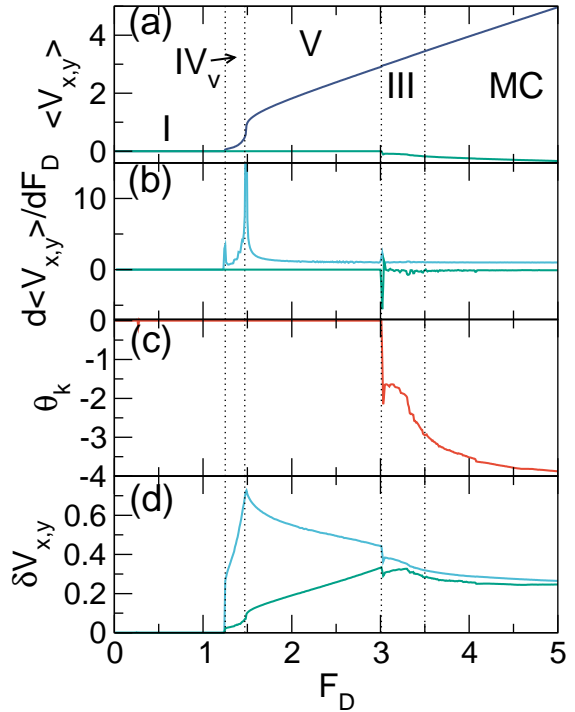}
\caption{
Transport signatures for a system 
with a finite Magnus term, where
$\alpha_m=0.075$, $\alpha_d=0.9972$,
the intrinsic Hall angle $\theta_H^{\rm int}=-4.29^\circ$,
$f=0.9536$, and $F_p=1.5$.
(a)
$\langle V_x\rangle$ (blue) and $\langle V_y\rangle$ (green)
vs $F_D$.
(b) The corresponding $d\langle V_x\rangle/dF_{D}$ (blue) and
$d\langle V_y\rangle/dF_D$ (green) vs $F_{D}$ curves.
(c) $\theta_{H} = \arctan(\langle V_{x}\rangle/\langle V_{y}\rangle)$
vs $F_D$.
(d) $\delta V_{x}$ (blue) and $\delta V_y$ (green) vs $F_D$.
The phases are I (pinned), 
IV$_v$ (vacancy flow), V (moving smectic),
III (disordered liquid), and MC (moving crystal).
}
\label{fig:9}
\end{figure}

\begin{figure*}
\includegraphics[width=0.75\textwidth]{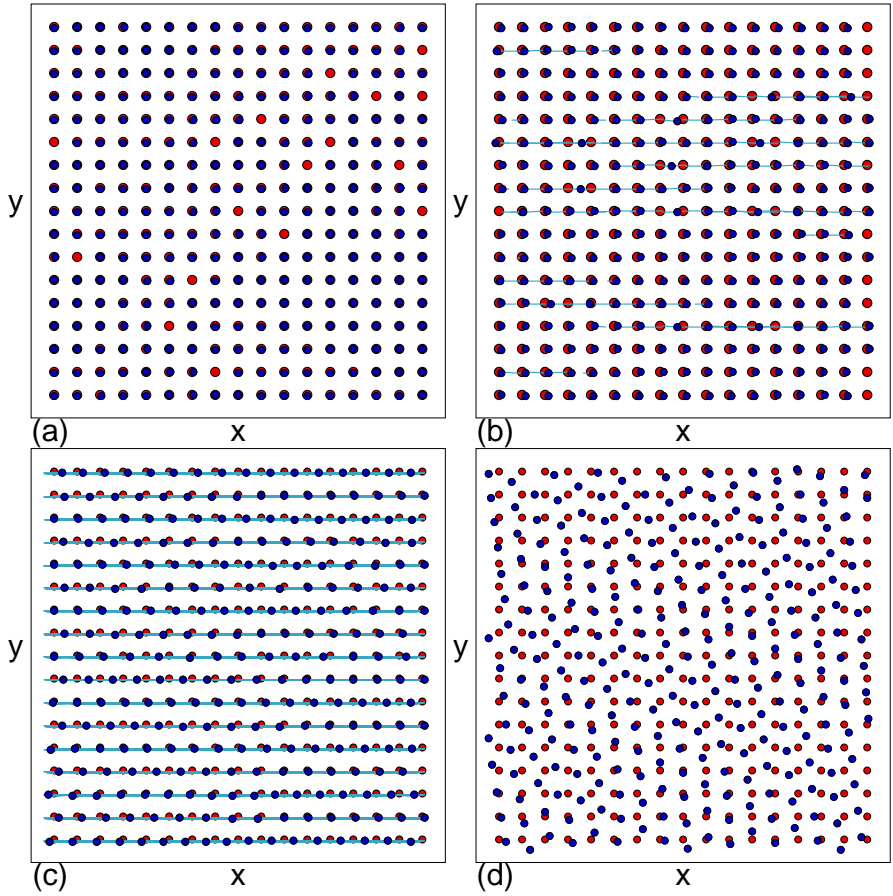}
\caption
{Images of particle (blue) and pin (red) positions along with particle
trajectories (light blue lines in (b) and (c))
for a portion of the system in Fig.~\ref{fig:9} with
$\alpha_m=0.075$, $\alpha_d=0.9972$, $\theta_H^{\rm int}=-4.29^\circ$, $F_p=1.5$,
and $f=0.9536$.
(a) Phase I at $F_{D} = 0.7$ showing a square lattice with pinned vacancies.
(b) Phase IV$_v$ at $F_{D} = 1.35$. 
(c) Phase V at $F_{D} = 1.7$ where all the particles are  moving.
(d) The MC phase at $F_{D} = 3.75$ where the particles are moving in
both the $x$ and $y$ directions (trajectories not shown).
}
\label{fig:10}
\end{figure*}

We next examine how these dynamic phases
evolve when a finite
Magnus force is introduced.
We first consider a filling of $f = 0.9536$, where there is a two step
depinning in which vacancies move above the first depinning transition
and the remaining vortices move above the second depinning transition.
In Fig.~\ref{fig:9}(a) we plot
$\langle V_{x}\rangle$ and $\langle V_{y}\rangle$
versus $F_{D}$
for a sample with
$\alpha_{m} = 0.075$ and $\alpha_d=0.9972$, where the intrinsic
Hall angle is $\theta^{\rm int}_{H} = -4.29^\circ$, 
and in Fig.~\ref{fig:9}(b)
we show the corresponding
$d\langle V_{x}\rangle/dF_{D}$ and $d\langle V_y\rangle/dF_D$
versus $F_{D}$.
The two step depinning transition persists, as indicated
by the pair of peaks in
$d\langle V_{x}\rangle/dF_{D}$ at $F_{D} = 1.28$ and
$F_D=1.5$.
The transverse velocity 
$\langle V_{y}\rangle = 0.0$ up to $F_{D} = 3.0$,
above which 
$\langle V_{y}\rangle$ becomes finite
and the particles start to move in the negative $y$-direction.
This transverse depinning transition is accompanied
by a dip in $d\langle V_{y}\rangle/dF_{D}$.
In Fig.~\ref{fig:9}(c) we plot the measured
Hall angle $\theta_{H} = \arctan(\langle V_{x}\rangle/\langle V_{y}\rangle)$,
which has a jump to a finite value near $F_{D} = 3.0$.
As $F_{D}$ increases, the magnitude of
$\theta_H$ increases and approaches the intrinsic value $\theta_H^{\rm int}$
at higher drives.
The plots of $\delta V_x$ and $\delta V_y$ versus $F_D$ in
Fig.~\ref{fig:9}(d)
indicate that in phase IV$_{v}$ and phase V,
$\delta V_{x} > \delta V_{y}$, 
while above the cusp at $F_{D}=3.0$,
$\delta V_{x}$ and $\delta V_{y}$ have nearly the 
same value.
Above the transverse depinning transition, when the particles begin
to move in the $y$ direction, the flow is in the disordered phase III
state, while
at higher drives the particles organize into
a moving crystal (MC) state.
When the Magnus term is finite,
the disordered flow phase III
always has a finite Hall angle.
Since the particles are moving at an angle, they do not lock to a
symmetry direction of the substrate, and only form the moving crystal
phase when the drive is high enough to cause the particles to float
above the substrate.

In Fig.~\ref{fig:10}(a) we illustrate the particle and pinning
site locations
for the system in Fig.~\ref{fig:9} at $F_{D} = 0.7$ where there is a 
square ground state containing pinned vacancies.  
At $F_D=1.35$ in phase IV$_a$, Fig.~\ref{fig:10}(b)
shows the particle trajectories along with the particle and pinning
site locations.
Here the vacancies move opposite to the driving
direction along 1D chains while the
individual particles in those chains move in the driving
direction one hop at a time, and the remaining particles are pinned.
In Fig.~10(c), at $F_D=1.7$ in phase V,
all of the particles are moving along the pinning rows
and exhibit a smectic ordering. 
Figure 10(d) shows only the particle and pin locations without
trajectories for the
moving crystal phase at $F_{D} = 3.75$, where the particles
form 
a triangular lattice that
translates at an angle with respect to the pinning lattice. 

\begin{figure}
\includegraphics[width=\columnwidth]{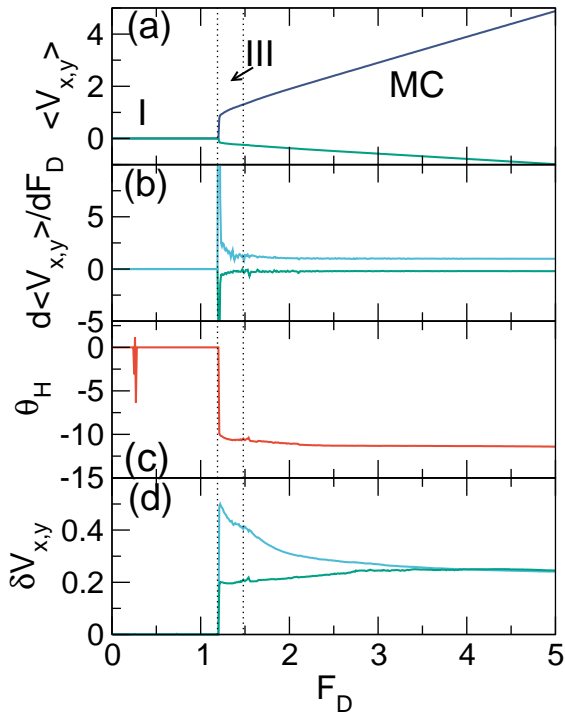}
\caption{
Transport signatures for system 
with a finite Magnus term, where
$\alpha_m=0.2$, $\alpha_d=0.9798$,
the intrinsic Hall angle $\theta_H^{\rm int}=-11.54^\circ$,
$f=0.9536$, and $F_p=1.5$.
(a)
$\langle V_x\rangle$ (blue) and $\langle V_y\rangle$ (green)
vs $F_D$.
(b) The corresponding $d\langle V_x\rangle/dF_{D}$ (blue) and
$d\langle V_y\rangle/dF_D$ (green) vs $F_{D}$ curves.
(c) $\theta_{H} = \arctan(\langle V_{x}\rangle/\langle V_{y}\rangle)$
vs $F_D$.
(d) $\delta V_{x}$ (blue) and $\delta V_y$ (green) vs $F_D$.
Here there is a transition directly from the pinned phase I to
the disordered flow phase III in which the
magnitude of the Hall angle increases with
increasing drive. At high drives the moving crystal (MC) phase appears.
}
\label{fig:11}
\end{figure}

When $\alpha_{m}$ increases, phases IV$_v$ and V are lost, as
shown in Fig.~\ref{fig:11} where we plot $\langle V_x\rangle$,
$\langle V_y\rangle$, $d\langle V_x\rangle/dF_D$,
$d\langle V_y\rangle/dF_D$, $\delta V_x$, and $\delta V_y$ versus
$F_D$ for a sample with
$\alpha_m = 0.2$, $\alpha_d=0.9798$, and 
$\theta^{\rm int}_{H} = -11.54^\circ$.
Here there is a single peak in
$d\langle V_x\rangle/dF_{D}$ and a single dip in $d\langle V_y\rangle/dF_D$
at the transition from the pinned phase
I to the fluctuating liquid phase III,
and $\theta_{H}$ rapidly approaches
the intrinsic value $\theta_{H}^{\rm int}$ just above the depinning
transition.
At higher drives, the fluctuations $\delta V_x$ and $\delta V_y$
parallel and transverse to the driving direction become nearly
identical in magnitude.

\begin{figure}
\includegraphics[width=\columnwidth]{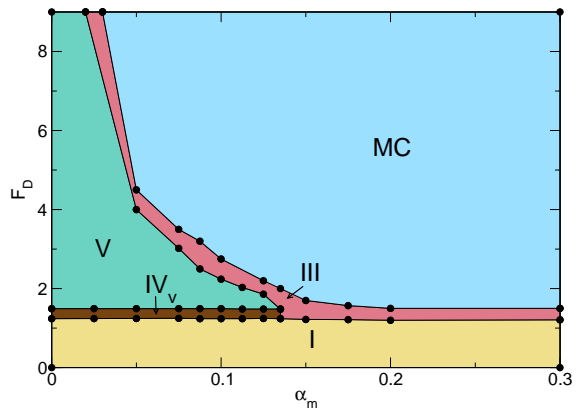}
\caption
{Dynamic phase diagram as a function of $F_D$ vs $\alpha_m$
constructed from the $\langle V_x\rangle$, $\langle V_y\rangle$,
$d\langle V_x\rangle$, $d\langle V_y\rangle$,
$\delta V_x$ and $\delta V_y$ curves
for the systems in Figs.~\ref{fig:9}, \ref{fig:10}, and \ref{fig:11}
with 
$f = 0.9536$,
showing phase I (pinned, yellow),
phase III (disordered flow, pink),
phase IV$_{v}$ (vacancy flow, dark brown),
phase V (moving smectic, green), and
the MC phase (moving crystal, light blue). 
}
\label{fig:12}
\end{figure}

From the features in the transport curves
for $f = 0.9536$,
we can construct a dynamic phase diagram
as a function of $F_{D}$ versus $\alpha_{m}$, as shown in
Fig.~\ref{fig:12}.
Here we find the pinned phase I,
the disordered or fluid flow phase III,
the 1D flow of vacancies in phase
IV$_{v}$,
the smectic flow phase V,
and the moving crystal phase MC.
In this case, the depinning threshold
is almost independent of $\alpha_{m}$.
In some studies, the velocity above depinning 
decreases with increasing $\alpha_{m}$
when the particle trajectories are bent around the pinning sites
\cite{Nagaosa13,Reichhardt15a}.
Since $\alpha_{m}$ can only modify the motion of a moving particle, if
the particle has settled into a pinned state, the depinning threshold
is insensitive to the value of $\alpha_m$.
It may be possible that if the drive were increased more rapidly from
zero, 
the depinning force could decrease with increasing
$\alpha_{m}$,
but we are working in the limit where the drive is incremented so slowly
that the depinning threshold has no rate dependence.
For $\alpha_{m} < 0.15$, the system depins from
phase I into phase IV$_v$ and then transitions into phase V.
At higher drives, there is a window of phase III separating phase V from
the moving crystal MC phase, which has a finite Hall angle.
The III-MC transition roughly follows
the line $1/\alpha_{m}$ for
$\alpha_{m} < 0.125$.
When $\alpha_{m} > 0.125$,
the system depins from phase I directly into phase III, which has
a finite Hall angle. 
In some cases, particularly for smaller $\alpha_{m}$,
we can observe some locking steps where the motion of the particles
locks to one of the symmetry directions of the pinning lattice.
Particularly strong directional locking occurs when
$\alpha_{m}/\alpha_{d} = 1.0$,
where $\theta_{H}^{\rm int}=45^\circ$ is aligned with one of the major
symmetry directions of the square pinning array.
Our results also show that there is
a range of Magnus terms over which the dynamical
behavior of the system remains essentially the same as that of an
overdamped system with $\alpha_m=0$.

\begin{figure}
\includegraphics[width=\columnwidth]{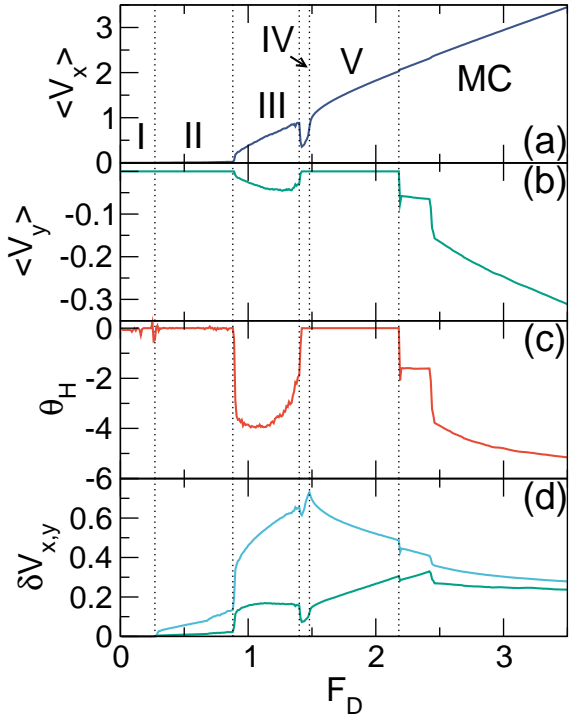}
\caption{
Transport signatures for a system 
with
$\alpha_m=0.1$, $\alpha_d=0.9945$,
$\theta_H^{\rm int}=-5.74^\circ$,
and $F_p=1.5$ at $f=1.005$.
(a)
$\langle V_x\rangle$
vs $F_D$.
(b)
$\langle V_y\rangle$
vs $F_D$.
(c) $\theta_{H} = \arctan(\langle V_{x}\rangle/\langle V_{y}\rangle)$
vs $F_D$.
(d) $\delta V_{x}$ (blue) and $\delta V_y$ (green) vs $F_D$.
The phases are: I (pinned), II (interstitial flow),
III (disordered flow with finite Hall angle),
IV (soliton flow), V (moving smectic),
and MC (moving crystal with a finite Hall angle). 
}
\label{fig:13}
\end{figure}

We next consider the impact of the Magnus term
on the behavior for $f > 1.0$. 
In Fig.~\ref{fig:13}
we plot $\langle V_{x}\rangle$, $\langle V_{y}\rangle$,
$\theta_{H}$, 
$\delta V_{x}$, and $\delta V_{y}$ versus $F_D$
for a system
with $f = 1.005$ at
$\alpha_{m} = 0.1$, $\alpha_d=0.9945$, and $\theta^{\rm int}_{H} = -5.74^\circ$. 
When the system depins from phase I into phase II, the
interstitial particles
move in the driving direction with a Hall angle of $\theta_H=0$.
In phase III,
$\langle V_{y}\rangle$ becomes finite and the magnitude of the
Hall angle increases up to $\lvert\theta_{H}\rvert=4.0^\circ$.
The transition to
phase IV is marked by a decrease of $\langle V_y\rangle$ to zero and
a drop in $\langle V_{x}\rangle$ that produces
negative differential conduction.
This can be viewed as an example of reentrant transverse pinning.
When $F_{D}/F_{p} > 1.0$, the system enters
phase V, and
near $F_{D} = 2.3$,
a moving crystal state emerges that is marked by
jumps in $\langle V_{y}\rangle$ and $\theta_{H}$
to finite values.
$\delta V_{x}$ increases with increasing $F_D$ in 
phases II and III,
dips during phase IV, and shows
another drop at the transition to the MC phase,
while $\delta V_{y}$ shows a drop in
phase IV and a jump up 
at the MC transition.
There is a locking step with finite fixed $\theta_H$ at the
beginning of the MC phase.
For smaller values of $\alpha_{m}$, we observe the same trends,
but the V-MC transition
shifts to higher values of $F_{D}$.  

\begin{figure}
\includegraphics[width=\columnwidth]{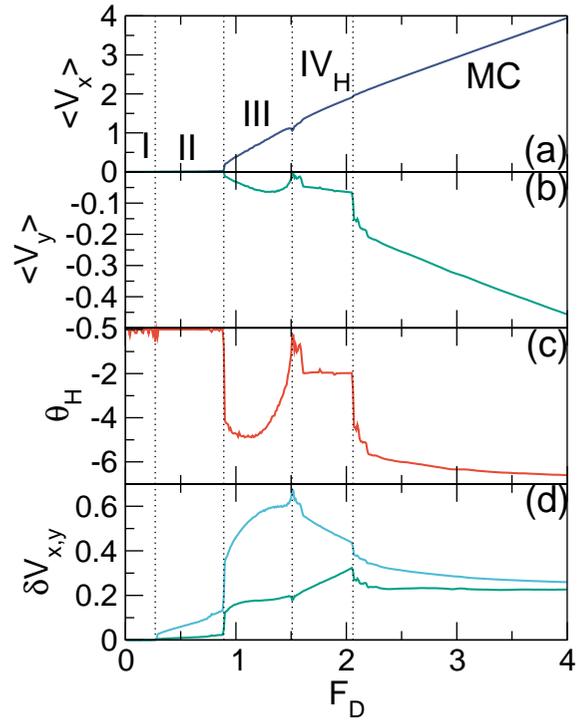}
\caption{
Transport signatures for a system 
with
$\alpha_m=0.12$, $\alpha_d=0.9928$,
$\theta_H^{\rm int}=-6.89^\circ$,
and $F_p=1.5$ at $f=1.005$.
(a)
$\langle V_x\rangle$
vs $F_D$.
(b)
$\langle V_y\rangle$
vs $F_D$.
(c) $\theta_{H} = \arctan(\langle V_{x}\rangle/\langle V_{y}\rangle)$
vs $F_D$.
(d) $\delta V_{x}$ (blue) and $\delta V_y$ (green) vs $F_D$.
The phases are I (pinned), II (interstitial flow), III (disordered flow
with a finite Hall angle), IV$_H$ (soliton flow with a finite Hall angle),
and MC (moving crystal with a finite Hall angle).
}
\label{fig:14}
\end{figure}

For $0.11 < \alpha_{m} < 0.3$,
an inversion of the velocity-force curves
occurs and
the features found in
$\langle V_{x}\rangle$ for $\alpha_m < 0.11$ now appear in
the $\langle V_{y}\rangle$ curves instead.
These include
negative differential conductivity
as well as an additional phase called IV$_{H}$ consisting of
solitons flowing with a finite Hall angle.
In Fig.~\ref{fig:14} we show 
$\langle V_{x}\rangle$, $\langle V_{y}\rangle$, $\theta_{H}$
$\delta V_{x}$, and $\delta V_y$ versus $F_D$ for a system
with $f = 1.005$ at
$\alpha_{m} = 0.12$,
$\alpha_d=0.9928$, and $\theta^{\rm int}_{H} = -6.89^\circ$.
We find a pinned phase I, an interstitial flow phase II with no Hall angle,
and a disordered flow phase III with a Hall angle close to
$\theta_H=-6^\circ$.
There is a region of negative differential conductivity in $\langle V_y\rangle$
near the III-IV$_H$ transition, and in phase IV$_H$, there is soliton
flow at a Hall angle of close to $\theta_H=-2.0^\circ$.
At high drives, there is a jump in the magnitude of $\langle V_y\rangle$ at
the onset of the moving crystal phase MC.
The III-IV$_H$ transition produces signatures in
$\theta_{H}$, $\delta V_{x}$, and $\delta V_y$.
In the previously observed phase IV, the solitons
are moving strictly along the $x$ direction with
$\theta_H=0^\circ$, whereas in phase IV$_H$, each soliton moves primarily along
the $x$ direction but makes periodic jumps in the $y$ direction, giving a
finite Hall angle.
As $\alpha_{m}$ increases,
the width of the IV$_H$ phase decreases
until the system transitions directly from phase III to phase MC.

\begin{figure}
\includegraphics[width=\columnwidth]{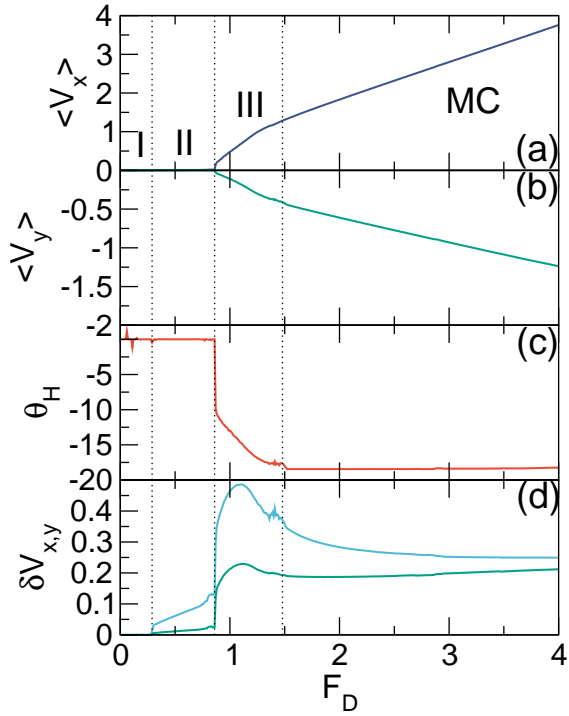}
        \caption{
Transport signatures for a system
with
$\alpha_m=0.3$, $\alpha_d=0.9539$,
$\theta_H^{\rm int}=-16.7^\circ$,
and $F_p=1.5$ at $f=1.005$.
(a)
$\langle V_x\rangle$
vs $F_D$.
(b)
$\langle V_y\rangle$
vs $F_D$.
(c) $\theta_{H} = \arctan(\langle V_{x}\rangle/\langle V_{y}\rangle)$
vs $F_D$.
(d) $\delta V_{x}$ (blue) and $\delta V_y$ (green) vs $F_D$.
The phases are: I (pinned), II (interstitial flow),
III (disordered flow with a finite Hall angle), and MC (moving
crystal with a finite Hall angle).
}
\label{fig:15}
\end{figure}

\begin{figure}
\includegraphics[width=\columnwidth]{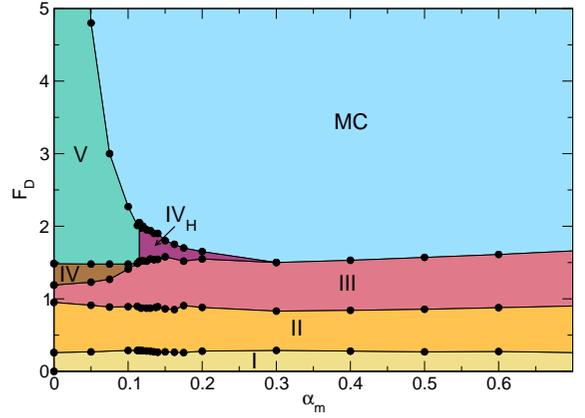}
\caption
{
Dynamic phase diagram as a function of
$F_{D}$ vs $\alpha_m$
for the system in Figs.~\ref{fig:13}, \ref{fig:14}, and \ref{fig:15}
with $f = 1.005$,
showing
phases I (pinned, yellow), II (interstitial flow with zero Hall angle,
light orange),
III (disordered flow with finite Hall angle, pink),
IV (soliton flow with zero Hall angle, medium brown), 
IV$_H$ (soliton flow with finite Hall angle, purple),  
V (moving smectic with zero Hall angle, green),
and MC (moving crystal with finite Hall angle, light blue).
}
\label{fig:16}
\end{figure}

In Fig.~\ref{fig:15} we plot $\langle V_{x}\rangle$,
$\langle V_{y}\rangle$, $\theta_{H}$, $\delta V_{x}$, and
$\delta V_y$ versus $F_D$
for a system with $f = 1.005$ at
$\alpha_{m} = 0.3$,  $\alpha_d=0.9539$, and $\theta^{\rm int}_{H} = -16.7^\circ$.
We observe the pinned phase I, the interstitial phase II with zero Hall
angle, the disordered flow phase III with a finite Hall angle that
increases in magnitude with increasing $F_D$, and a moving crystal with
a Hall angle close to the intrinsic value.
Phase IV$_H$ is now missing, and $\delta V_{x}$
shows a peak near the III-MC transition.

We use the features in the velocity-force, 
$\theta_{H}$, $\delta V_{x}$, and $\delta V_y$
curves to construct the dynamical phase diagram for
$f = 1.005$
as a function of $F_{D}$ versus $\alpha_{m}$,
shown in Fig.~\ref{fig:16}.
Phases I, II, and III are fairly insensitive to the value of $\alpha_m$,
although the disordered flow phase III now has a finite Hall angle.
The soliton flow phase IV and
smectic flow phase V, both with zero Hall angle, vanish above
$\alpha_m=0.11$. There is a window of phase IV$_H$, or soliton flow with
a finite Hall angle, separating phase III from the moving crystal MC phase
with finite Hall angle for $0.11<\alpha_m<0.3$.

\begin{figure}
\includegraphics[width=\columnwidth]{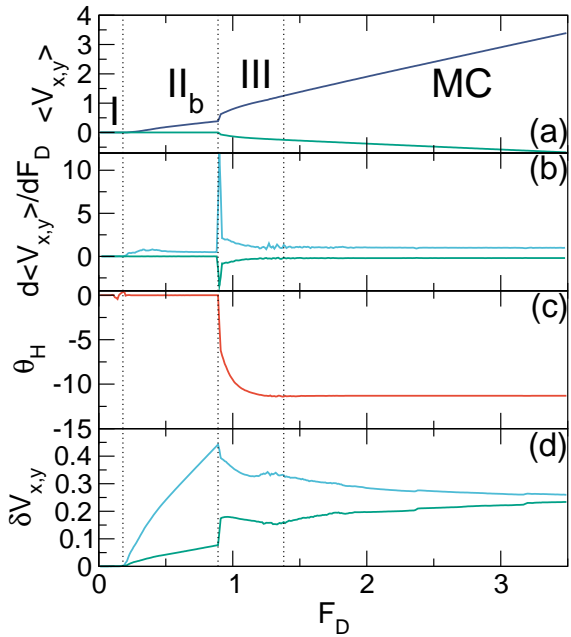}
        \caption{
Transport signatures for a system 
with $\alpha_m=0.2$, $\alpha_d=0.9798$, $\theta_H^{\rm int}=-11.54^\circ$,
$F_p=1.5$, and
$f=1.85$.
(a)
$\langle V_x\rangle$ (blue) and $\langle V_y\rangle$ (green)
vs $F_D$.
(b) The corresponding $d\langle V_x\rangle/dF_{D}$ (blue) and
$d\langle V_y\rangle/dF_D$ (green) vs $F_{D}$ curves.
(c) $\theta_{H} = \arctan(\langle V_{x}\rangle/\langle V_{y}\rangle)$
vs $F_D$.
(d) $\delta V_{x}$ (blue) and $\delta V_y$ (green) vs $F_D$.
Here we observe pinned phase I, interstitial flow phase II with zero Hall angle,
disordered flow phase III with a finite Hall angle, 
and the moving crystal MC phase with a finite Hall angle.
}
\label{fig:17}
\end{figure}

\begin{figure}
\includegraphics[width=\columnwidth]{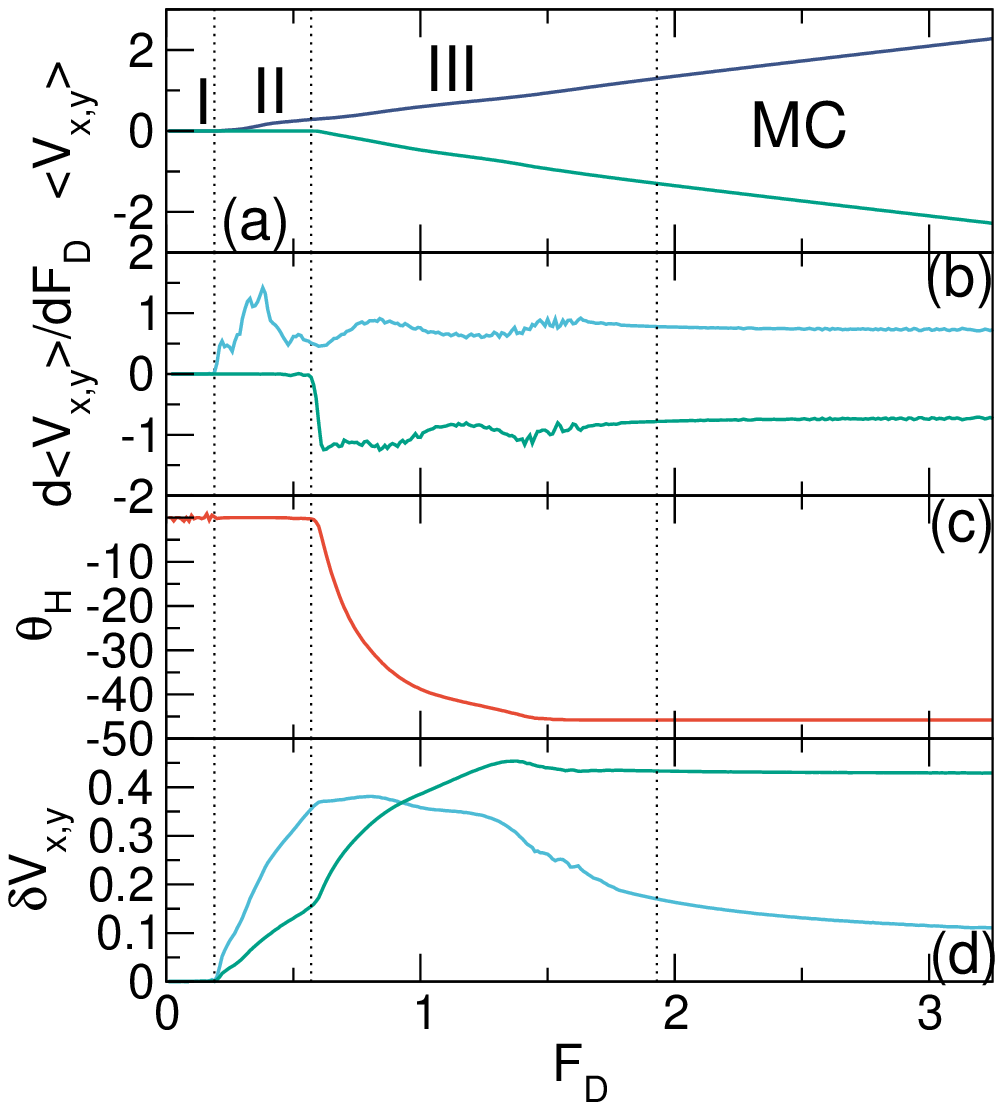}
        \caption{
Transport signatures for a system 
with $\alpha_m=1.0$, $\alpha_d=1.0$, $\theta_H^{\rm int}=-45^\circ$,
$F_p=1.5$, and
$f=1.005$.
(a)
$\langle V_x\rangle$ (blue) and $\langle V_y\rangle$ (green)
vs $F_D$.
(b) The corresponding $d\langle V_x\rangle/dF_{D}$ (blue) and
$d\langle V_y\rangle/dF_D$ (green) vs $F_{D}$ curves.
(c) $\theta_{H} = \arctan(\langle V_{x}\rangle/\langle V_{y}\rangle)$
vs $F_D$.
(d) $\delta V_{x}$ (blue) and $\delta V_y$ (green) vs $F_D$.
We observe pinned phase I, interstitial flow phase II,
disordered flow phase III with a finite Hall angle, and a moving
crystal phase MC with a finite Hall angle.
}
\label{fig:18}
\end{figure}

\begin{figure}
\includegraphics[width=\columnwidth]{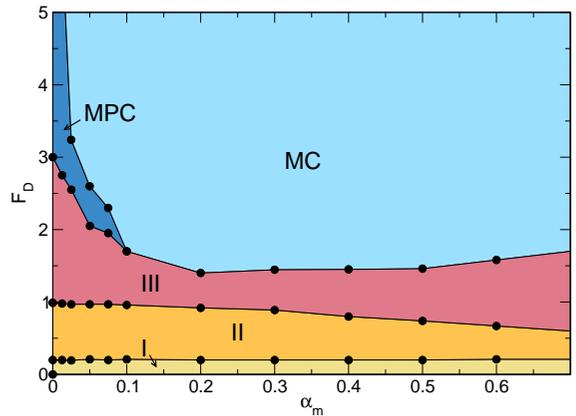}
\caption
{Dynamic phase diagram as a function of $F_D$ vs $\alpha_m$
for the system in Figs.~\ref{fig:17} and \ref{fig:18}
with $f = 1.85$,
showing
phases I (pinned, yellow), II (interstitial flow with zero Hall angle, light
orange),
III (disordered flow with finite Hall angle, pink),
MPC (moving polycrystalline, medium blue),
and MC (moving crystal, light blue).
}
\label{fig:19}
\end{figure}

We have also considered the evolution of the
different phases at higher fillings. 
In Fig.~\ref{fig:17} we plot the transport curves
for the system from Fig.~\ref{fig:1} at $f = 1.85$ with
$\alpha_{m} = 0.2$, $\alpha_d=0.9798$, and $\theta_H = -11.54^\circ$.
There is a region of phase II$_b$ flow with motion in both the interstitial
regions and along the pinning rows at zero Hall angle.
This is followed by a transition to the disordered flow phase III
and then to the moving crystal MC phase, both with a
finite Hall angle, and both detected using
peaks in $d\langle V_x\rangle/dF_D$, $d\langle V_y\rangle/dF_D$,
$\delta V_{x}$, and $\delta V_y$. 

In Fig.~\ref{fig:18}
we plot $\langle V_x\rangle$, $\langle V_y\rangle$,
$d\langle V_x\rangle/dF_{D}$, $d\langle V_y\rangle/dF_D$,
$\theta_H$, $\delta V_{x}$, and $\delta V_y$ versus $F_D$
for the same system in Fig.~\ref{fig:17} but at $\alpha_{m} = 1.0$,
$\alpha_d=1.0$, and
$\theta^{\rm int}_{H} = -45^\circ$.
In this case, at low drives we find phase I along with the
$\theta_{H} = 0$ interstitial flow phase II, which contains
multiple peaks in $d\langle V_x\rangle/dF_{D}$.
For $F_{D} > 1.5$, both $d\langle V_x\rangle/dF_D$ and
$d\langle V_y\rangle/dF_D$ reach constant values, and $\theta_H$ locks
to $\theta_H=-45^\circ$ in the moving crystal phase.

In Fig.~\ref{fig:19} we show the dynamic phase diagram as
a function of $F_D$ versus $\alpha_m$
for the system in Figs.~\ref{fig:17} and \ref{fig:18}
at $f = 1.85$.
The pinned phase I and interstitial flow phase II vary little as
$\alpha_m$ changes. The disordered flow phase III decreases in extent with
increasing $\alpha_m$ before saturating in width near $\alpha_m=0.2$.
The moving polycrystalline phase MPC appears at higher drives only
when $\alpha_m<0.1$, and is replaced by the moving crystal MC phase at
higher $\alpha_m$.

\section{Summary}
We have investigated the dynamic
phases of superconducting vortices and Magnus particles such as
skyrmions interacting with a square periodic pinning array.
For the superconducting vortex case where the Magnus force is zero and
there is only a damping term,
we map
the different phases as a function of filling up to the
second matching filling.
Just above the first matching filling, pinned vortices
coexist with a small number of interstitial vortices in the regions
between the pinning sites, and
as a function of drive, a series of phases appear including pinned,
interstitial flow, disordered fluid flow, soliton flow, and smectic
flow.
At the transition from fluid to soliton flow,
there is strong negative differential conductivity. 
We show that the differential conductivity curves
and the variance in the velocity fluctuations provide complementary
information to the previously considered velocity-force curves, and
make it possible to identify several previously overlooked phases.
For fillings just below the first matching filling,
there is a flow of anti-kinks or solitons along
with a double peak in the differential conductivity curves.
When a small Magnus force is present,
we find that these same phases
can occur but that the fluid phases
exhibit a finite Hall angle while the
interstitial, soliton, and smectic phases do not.
We also find a reentrant
Hall effect in which the Hall angle is finite at lower drives in the
fluid phase but drops back to zero at higher drives when the system
enters a soliton flow state.
For somewhat higher Magnus forces,
we find that the transverse velocity-force curves exhibit
the same features found in the longitudinal velocity-force curves
at zero Magnus force. Specifically, regions of negative
differential conductivity appear,
and a new soliton phase emerges that has a finite Hall angle.
We map out the
different phases a function of filling and
the Magnus force component.
Our results should be relevant for
superconducting systems in which the vortices have a finite Magnus
term and for
skyrmion systems in the regime of smaller
skyrmion Hall effects.
Additionally, our
work indicates that kinks, antikinks,
or interstitial flow of finite Magnus term particles such as
skyrmions on 2D pinning arrays
are associated with zero or reduced Hall angles,
which could be useful for device creation.

%\acknowledgements
\noindent{\textbf{\textsf{Data availability}}}\\
The datasets generated during and/or analyzed during
the current study are available from the corresponding
author upon reasonable request.

\noindent{\textbf{\textsf{Acknowledgements}}}
This work was supported by the US Department of Energy through
the Los Alamos National Laboratory and Research Foundation-Flanders (FWO).  
Los Alamos National Laboratory is
operated by Triad National Security, LLC, for the National Nuclear Security
Administration of the U. S. Department of Energy (Contract No. 892333218NCA000001).

\noindent{\textbf{\textsf{Competing Interests}}}\\
The authors declare no competing interests.

\noindent{\textbf{\textsf{Author contributions}}}\\
All authors contributed equally to this work.

\bibliographystyle{sn-mathphys}
%% BioMed_Central_Bib_Style_v1.01

\end{document}